\documentclass[a4paper,11pt]{article}
\usepackage{pos}
\renewcommand{\logo}{\relax}

\title{LUXE: A new experiment to study non-perturbative QED \\ in electron-laser and photon-laser collisions}
\ShortTitle{LUXE: A new experiment to study non-perturbative QED}

\manuallySeparateAuthors
\author*[a]{Shan Huang}
\author{ for the LUXE Collaboration}
\affiliation[a]{Tel Aviv University,\\
  Tel Aviv-Yafo 6997801, Israel}
\emailAdd{shan.huang@desy.de}

\abstract{
LUXE (Laser Und XFEL Experiment) is a new experiment in planning at DESY Hamburg using the electron beam of the European XFEL.
LUXE is intended to study collisions between a high-intensity optical laser and 16.5 GeV electrons from the XFEL electron beam, as well as collisions between the optical laser and high-energy secondary photons.
The physics objective of LUXE are processes of quantum electrodynamics (QED) at the strong-field frontier, where the electromagnetic field of the laser is above the Schwinger limit.
In this regime, QED is non-perturbative.
This manifests itself in the creation of physical electron-positron pairs from the QED vacuum, similar to Hawking radiation from black holes.
LUXE intends to measure the positron production rate in an unprecedented laser intensity regime.
An overview of the LUXE experimental setup and its challenges will be given, followed by a discussion of the expected physics reach in the context of testing QED in the non-perturbative regime.
}

\FullConference{%
  *** Particles and Nuclei International Conference - PANIC2021 ***\\
  *** 5 - 10 September, 2021 ***\\
  *** Online ***
}

\begin{document}
\maketitle

\newcommand{\units}{\ \mathrm} 

\section{Introduction}
Quantum electrodynamics is an important part of the standard model that forms a cornerstone of modern physics.
In the perturbative regime, QED gave one of the most precise predictions comparing to the experimental results, such as the electron's $g$ factor \cite{Hanneke_2008}.
However, in strong-field regime, the perturbative method's breakdown takes place when field-induced non-linear processes become dominating, and raises questions to be verified via experiment.
This task of verification motivates both the particle physicists working at GeV scales and atomic physicists at eV scales around the world to conduct ground-breaking experiments using high-power laser techniques \cite{E144_1999, Cole_2018,AstraGemini_2018, ELINP_2017, E320_2019}.

It has been predicted that the vacuum will ``boil'', indicating photons spontaneously conversion into real electron-positron pairs at a critical field, quoted as the Schwinger limit \cite{Sauter_1931, EulerHeisenberg_1936, Schwinger_1951}, where the energy density equals the electron's mass ($m_e$) confined in the cube of its reduced Compton radius.
This limit is equivalent to an electric field
\vspace{-0.35 cm}
\begin{equation}
{\cal E}_\mathrm{cr} = \frac{c^3}{\hbar e} \sqrt{m_e^4}
	= 1.32 \times {10}^{16} \units{V\,{cm}^{-1}}.
\vspace{-0.35 cm}
\end{equation}

The physics in this regime is relevant to the astrophysical phenomena near black holes \cite{Ruffini_2010} where the strong field is also present.
Moreover, the physics of the ``transition'' regime, where the perturbative processes turn into the fully non-perturbative ones, also has some appealing phenomena like the vacuum polarisation and birefringence along with the quantum radiation reaction \cite{Blackburn_2020}.

The ``Laser und XFEL Experiment'' or LUXE, proposed by DESY Hamburg and the European XFEL (XFEL.EU), with over 88 researchers from 29 institutes joining in, has a goal to precisely measure the physics at the non-perturbative QED regime, including
(1) the interactions of real photons with electrons and photons at the critical field,
(2) the electron-photon and photon-photon interactions in the transition regime, as well as
(3) the possible creation of particles beyond the standard model (such as axion-like-particles) in strong fields.
The experiment will be conducted by observing the collisions between a TW or sub-PW optical laser and a high-quality GeV electron or converted photon beam from the XFEL.EU.
Such collisions will allow us to study the non-perturbative physics near or even above the Schwinger limit.
LUXE will be one of the first experiments to reach the uncharted regime with its first data to be taken in 2024.

\begin{figure}[!b] 
   \centering
   \includegraphics[height=4.5 cm]{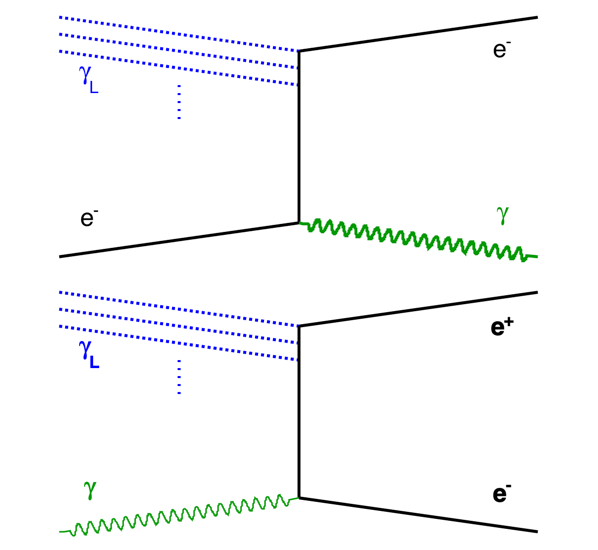} \quad
   \includegraphics[height=4.2 cm]{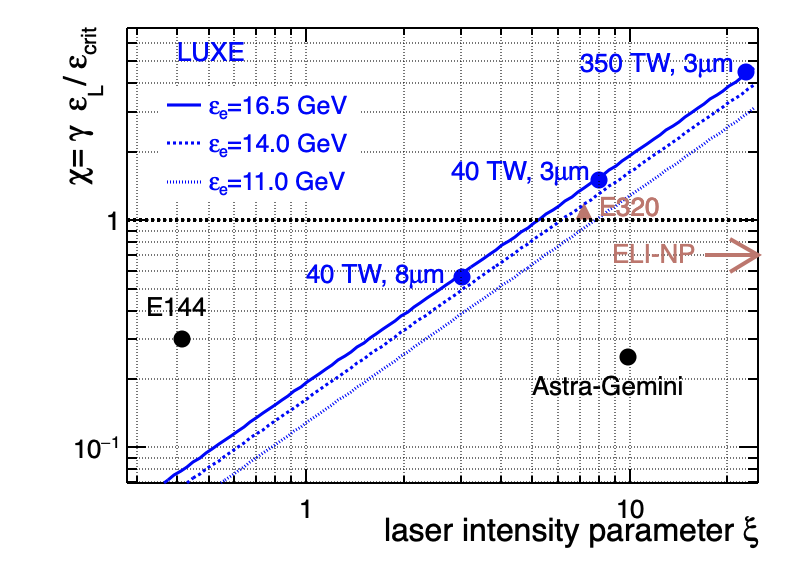}
   \caption{Left: Feynman diagrams of (up) nonlinear Compton scattering and (down) Breit--Wheeler process. Right: Quantum parameter $\chi$ vs the intensity parameter $\xi$ for LUXE \cite{LUXECDR_2021} and a selection of experiments and facilities. E320 and ELI-NP are not yet operating.}
   \label{fig:feyn}
   \label{fig:chi_xi}
\end{figure}

\section{Theory and Simulation}
There are three processes of interest at the interaction point (IP):
(1) the non-linear Compton scattering,
(2) the non-linear Breit--Wheeler process [Fig. \ref{fig:feyn} (left)], and
(3) the non-linear trident process (dominated by the two-step process in LUXE parameter space [Fig. \ref{fig:chi_xi} (right)]).
The first and the third processes are expected in the electron-laser setup, while the second is to be observed in the gamma-laser setup.
The strong-field QED theory for the three processes lies in a parameter space with two dimensionless variables.
One is the classical non-linearity parameter, or often referred as the intensity parameter
\vspace{-0.2 cm}
\begin{equation}
\xi = \frac{e {\cal E}_\mathrm{L}}{m_e c \omega_\mathrm{L}} \propto \sqrt{I_\mathrm{L}},
\vspace{-0.2 cm}
\end{equation}
where ${\cal E}_\mathrm{L}$, $\omega_\mathrm{L}$, and $I_\mathrm{L}$ are the laser's electric field, angular frequency, and intensity, respectively.
This Lorentz-invariant parameter indicates a transition from the perturbative process to the non-linear interaction.
In the LUXE specifications, $\xi$ can reach an order of magnitude of $10^1$. However, reaching the critical field requires $\xi_\mathrm{cr} = 3.29 \times 10^5$ for 1.5 eV photons.
A Lorentz boost to the centre-of-mass reference frame is used to enhance the field strength to reach the fully non-perturbative regime while the physical laws unchange.
This boost is described by a quantum parameter

\vspace{-0.25 cm}
\begin{equation}
\chi = \frac{|p_\mu k^\mu_\mathrm{L}|}{m_e^2 c^2} \xi
	= (1+\beta \cos{\theta}) \frac{\hbar \omega_\mathrm{L} E}{m_e^2 c^4} \xi,
\vspace{-0.15 cm}
\end{equation}
in which $p$ stands for the four-momentum of the probe electron or photon, and $k$ for the one of the laser photon.
The variable $\beta$ is the probe particle's relativistic parameter while $\theta$ is the collision angle (0\textdegree\ for head-on collision).
With about $\xi = 5$, one can reach $\chi = 1$ in the LUXE setup.

\begin{figure}[!b] 
   \centering
    \vspace{-0.2 cm}
   \includegraphics[height=4.2 cm]{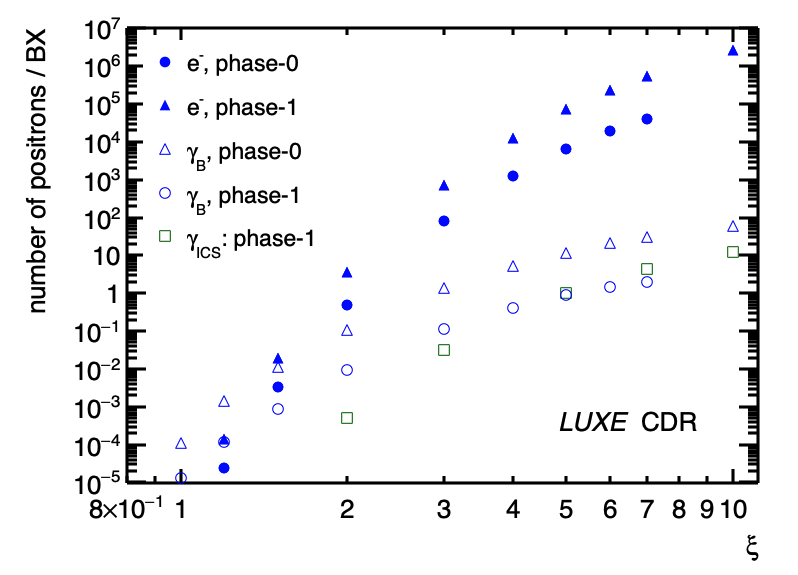} \quad
   \includegraphics[height=4.5 cm]{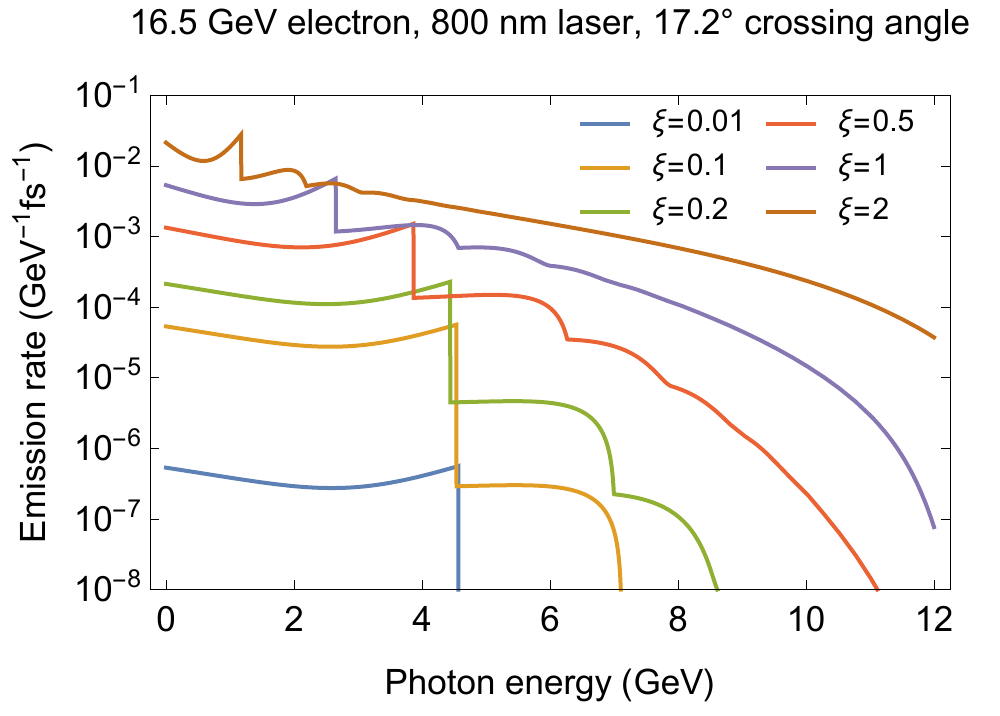}
   \caption{Left: number of positrons per beam crossing produced in e/$\gamma$-laser setups with various lasers and constant $\xi$ values. Right: Compton photon emission rate spectrum with a selection of constant $\xi$ values. \cite{LUXECDR_2021}}
   \label{fig:rate}
\end{figure}

The transition from the fully perturbative to the fully non-perturbative case can be shown and tested in the change of scaling rules.
In the low-field regime, the production rate merely depends on the number of particles reaching the IP per unit time.
With the increase of the laser intensity, the rate contributed by multi-photon processes will grow, which has been verified by the E144 experiment \cite{E144_1999}.
At high field regime ($\xi \gg 1$), the rate will become solely depending on various forms of $\chi$ functions in different areas.
This kind of transition can also be seen in the energy spectrum's edge shift in Compton scattering.
Furthermore, since the theories are based on plane-wave assumptions and limited in describing a real world highly focused laser,
a strong-field QED Monte Carlo simulation code, \textsc{Ptarmigan}, has been custom-built and used for LUXE \cite{Ptarmigan}.
\textsc{Ptarmigan} is based on the local monochromatic approximation (LMA), which is more suitable for the low $\xi$ regime.
As shown in Fig. \ref{fig:rate}, the MC simulations give a wide range of over 7 orders of magnitude in LUXE parameter-space for detectors to cover.

\section{Experimental Setup}
\subsection{Beams}
The high power laser beam for LUXE has two options. In Phase 0, the ``JETI-40'', a 800-nm Ti:Sapphire laser with 40 TW power and 30 fs pulse, will be used. And
in Phase 1, the beam will be upgraded to a commercial 350 TW laser.
Focused in 3 {\textmu m} spot, those lasers can reach their peak $\xi$ at 7.9 and 23.6, respectively.
The lasers will operate at 1 Hz repetition rate and use a high precision laser diagnose system to ensure less than 5\% peak and 1\% shot-to-shot intensity uncertainties.

The electron beam used by LUXE will come through the XFEL.EU 1.9 km linear accelerator.
Every electron train contains 2700 bunches of electron beams, and one bunch of the 16.5 GeV beam will be guided into the LUXE shaft.
Each bunch has $1.5\times 10^9$ electrons at 10 Hz repetition rate that allows the data taking at 9 beam-only runs and 1 beam-and-laser run mode for in situ background study.
The electron beam can directly interact with the laser, or be converted to GeV photons via bremsstrahlung.
The collision angle $\theta$ is set at 17.2{\textdegree}.
After the boost, $\chi$ can reach a value of 1.5 in Phase 0 and 4.5 in Phase 1.

\begin{figure}[!h] 
   \centering
   \includegraphics[height=4.2 cm]{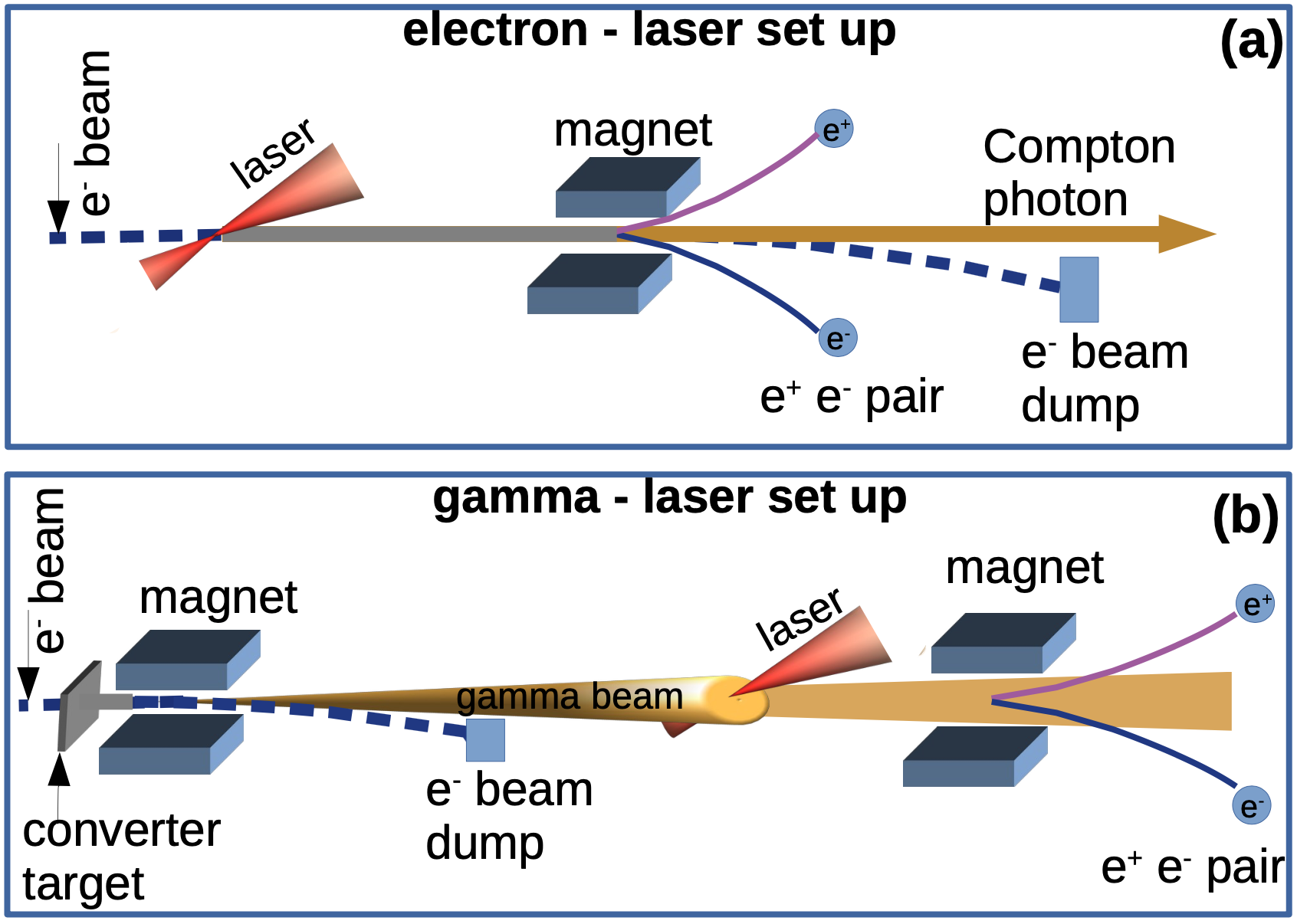} \quad
   \includegraphics[height=4.2 cm]{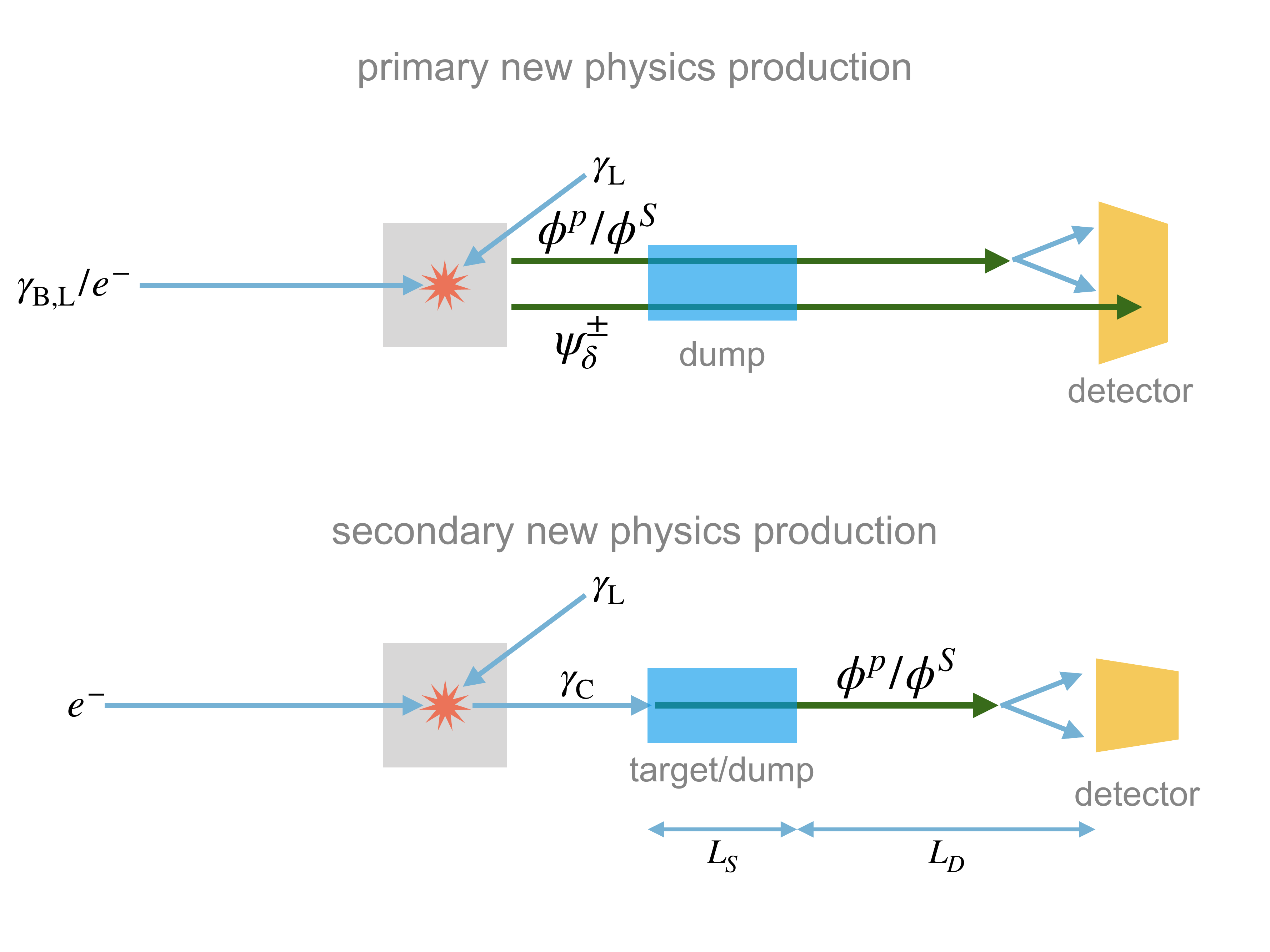}
   \caption{Left: schematic experimental layouts for the (a) electron-laser and (b) gamma-laser setups. \newline Right: schematic design of the LUXE new physics search setup at (top) the IP and (bottom) the optical dump. Symbols $\phi^{P}$/$\phi^{S}$ stand for ALPs and other scalar particles that interact with photon.
\cite{LUXECDR_2021}}
   \label{fig:setup}
\vspace{-0.55 cm}
\end{figure}

\subsection{Detectors}
As shown in Fig. \ref{fig:setup} (left), after the collision, the signal along with the rest of the probe beam will go through a magnetic dipole and then be separated.
The signal electrons will be detected by Argon-filled Cherenkov detectors for high-multiplicity flux and a scintillator screen with high position resolution.
The signal positrons, with relatively lower multiplicity, will be picked up by four-layer ALPIDE tracking detectors and a high-granularity electromagnetic calorimeter.
The high-flux GeV photon beam, however, will go straight along the beam-line to the gamma spectrometer, gamma profiler, and gamma flux monitor.
To suppress the background from the probe and improve the data collecting efficiency, a series of geometry simulations are performed using a \textsc{Geant4} code.
The details of the detecting system can be found in these proceedings by Y. Benhammou \cite{Yan}.

In both the electron-laser and gamma-laser setups, a strong field will be presented at the IP, which increases the possibility of the creation of axion-like-particles (ALPs) via Primakoff process.
While the Compton and bremsstrahlung photons are stopped at the beam dump, the ALPs can pass through it and arrive in a clean room behind the dump and make a light-shinning-through-wall experiment feasible, shown in Fig. \ref{fig:setup} (right).
Moreover, during the hitting of the high-flux photons with the dump, ALPs can be created locally.
The details of the new physics in LUXE can be found in the proceedings by A. Santra \cite{Santra} and the LUXE-NPOD article \cite{Bai_2021}.

\section{Conclusion}
LUXE will be one of the first experiments to explore QED in the uncharted strong-field frontier.
With a close cooperation between the high-energy particle physics and high-power laser physics communities, the physics of collisions between a high-quality electron/photon beam and a highly stable laser has been studied in theory and simulation, including the nonlinear Compton scattering and nonlinear pair creation by real photons, as well as the verification of axion-like-particles and other scalar particles that interact with photon.
The experimental setup that can be adapted to a large dynamic range has been designed and is going to be tested and installed in the shaft.
The first data taking of LUXE is scheduled at 2024.

\section*{Acknowledgments}
This work was partly supported by the German-Israel Foundation (GIF).


\providecommand{\href}[2]{#2}\begingroup\raggedright\endgroup

\end{document}